# SPHEREx Re-Observation of Interstellar Object 3I/ATLAS in December 2025: Detection of Increased Post-Perihelion Activity, Refractory Coma Dust, & New Coma Gas Species

**C.M. Lisse,** Johns Hopkins University Applied Physics Laboratory, Planetary Exploration Group, Space Department, 11100 Johns Hopkins Road, Laurel, MD 20723, USA

**Y.P. Bach,** Korea Astronomy and Space Science Institute, 776 Daedeok-daero, Lee Won Chul Hall 301, Yuseong-gu, Daejeon 34055, Republic of Korea

**S.A. Bryan,** Arizona State University, 650 E Tyler Mall, Tempe, AZ 85281, USA

**P. M. Korngut,** Division of Physics, Mathematics and Astronomy, California Institute of Technology, Pasadena, CA 91125, USA

**B.P. Crill,** California Institute of Technology/Jet Propulsion Lab, 4800 Oak Grove Drive, Pasadena, CA 91009, USA

**A.J. Cukierman,** Division of Physics, Mathematics and Astronomy, California Institute of Technology, Pasadena, CA 91125, USA

**O. Doré,** California Institute of Technology/Jet Propulsion Lab, 4800 Oak Grove Drive, Pasadena, CA 91009, USA

**A. Cooray,** Department of Physics & Astronomy, University of California Irvine, Irvine, CA 92697, USA

**B. Fabinsky,** California Institute of Technology/Jet Propulsion Lab, 4800 Oak Grove Drive, Pasadena, CA 91009, USA

**A.L. Faisst,** Infrared Processing and Analysis Center, Caltech, 770 South Wilson Street, Pasadena, CA 91125, USA

**H. Hui,** Department of Physics, California Institute of Technology, 1200 E. California Boulevard, Pasadena, CA 91125, USA

**G.J. Melnick,** Center for Astrophysics, Harvard & Smithsonian, 60 Garden Street, Cambridge, MA 02138, USA

**C.H. Nguyen,** Department of Physics, California Institute of Technology, 1200 E. California Boulevard, Pasadena, CA 91125, USA

**Z. Rustamkulov,** Infrared Processing and Analysis Center, Caltech, 770 South Wilson Street, Pasadena, CA 91125, USA

**V. Tolls,** Center for Astrophysics, Harvard & Smithsonian, 60 Garden Street, Cambridge, MA 02138, USA

**P.Y. Wang,** Arizona State University, 650 E Tyler Mall, Tempe, AZ 85281, USA

**M.W. Werner,** California Institute of Technology/Jet Propulsion Lab, 4800 Oak Grove Drive, Pasadena, CA 91009, USA

**and the SPHEREx Science Team**

**Abstract.** In December-2025, SPHEREx re-observed 3I/ATLAS post-perihelion, finding a much more active object compared to August-2025 SPHEREx pre-perihelion observations, with marked evidence for development into an cometary body fully sublimating all its ices. The new imaging spectrophotometry was dominated by spatially resolved features due to light scattered by dust, along with thermal emission, plus gas-line-emissions from CN(~0.93um), $H_2O$(~2.7um), organic C-H(3.2–3.6um), $CO_2$(4.25–4.27um), and CO(~4.7um). The $CO_2$ gas-coma continued to be extended out to ~3'radius. The continuum spectral signature of $H_2O$-ice absorption had mostly disappeared, replaced by scattered-light plus thermal-emission from organo-silicaceous dust grains while the $H_2O$ gas-emission is 40x higher. The CO- and $CO_2$-gas comae were circularly symmetric, while the other comae appear morphologically similar to the dust-coma with its pear-shaped, solar-pointing, large icy dust grains dust tail. The new appearance of CN and C-H features suggests that these carbon-rich ice species were contained either in $H_2O$ phases or were trapped under them.

**1. SPHEREx Observations**. From 08-to-15-Dec-2025, the NASA SPHEREx mission (Bock+2025) re-observed interstellar object 3I/ATLAS at $r_h$=2.0-2.2au, Delta = 1.845-1.809au using R=40-to-130, 0.75-5.0um spectrophotometric imaging in 102 discrete spectral channels. The observations consisted of 104 special pointings with $i_{Time}$=115sec and 3I/ATLAS centered on SPHEREx's LVF-patterned detectors. As for SPHEREx pre-perihelion August-2025 3I observations (Lisse+2025a,b), photometry was obtained by locating 3I frames, placing a 2-pixel radius aperture at 3I's location, and subtracting the background counts from a surrounding annulus.

**2. Spectral Results**. The aggregate SPHEREx imaging spectrophotometry results are shown in Figure 1. The spectral continuum is now similar to SPHEREx measurements of zodiacal dust emission, dominated by scattered sunlight from 0.75–3.0um, and by dust thermal emission at 4-5um. A new feature is detected at 0.925um, coincident with a strong emission line of CN gas and consistent with recent reports of optical CN emission at 0.39um. Using updated gas production rates for 3I in August (Lisse+2025b) we find that the $H_2O$-gas feature, barely detected in August, is now ~40x stronger at $Q_{gas,H2O}$~1.4+/-0.28(1$\sigma$)x$10^{28}$ molec/sec, suggesting water is now fully sublimating. New emission detected strongly in bands from 3.2–3.6um denotes the presence of abundant gaseous $CH_3OH$, $H_2CO$, $CH_4$, and/or $C_2H_6$ (hereafter C-H) at $Q_{gas,organics}$~2.0+/-0.5(1$\sigma$)x$10^{27}$ molec/sec in the coma. The $CO_2$ emitted flux is only about 2x larger than in August, at $Q_{gas,CO2}$~3.0+/-0.45(1$\sigma$)x$10^{27}$ which suggests that this species was near-fully active pre-perihelion. The CO emitted flux has markedly increased ~80x since August, at $Q_{gas,CO}$~7.6+/-1.5(1σ)x$10^{27}$ meaning that the CO/$CO_2$ abundance ratio = 2.5 has also increased ~40-fold to become normal for a CO-dominated comet (Harrington-Pinto+2022). Along with the greatly enhanced $H_2O$ flux and new C-H species emission, this implies that a new ice reservoir is now active along with the one continuing to supporting the giant $CO_2$-coma.

**3. Imaging Results.** The gas-comae detected by SPHEREx were all resolved, extending from 1'-to-3' in radius (Figure 1), and all except the CN and C-H organics comae are markedly round with respect to the Sun and orbital velocity directions. By contrast, the SPHEREx continuum dust and organics images are markedly pear-shaped, with the "pear-stem" pointing sunward. The very different morphologies suggest that CN, C-H, and $H_2O$ are sourced from the dust, while the $CO_2$, and CO-gas is from a symmetric region centered on the nucleus. No obvious jet or small-particle dominated anti-solar tail structures were found.

**4. Temporal Variability**. Continuum SPHEREx geometry-corrected flux measurements of 3I at 1.0-1.5μm were corrected for the solar spectrum and plotted against time. We see no evidence for more than 20% variability over 08–to-15-December-2025.

**5. Discussion/Interpretation.** The December-2025 observations are consistent with a comet that is now fully active, sublimating even water ice on 08-Dec-2025 at $Q_{gas,H2O}$~1.4x$10^{28}$ molec/sec into a 16,500 km aperture at 0.60 km/sec. The apparent composition of the material making up 3I/ATLAS has changed, however. The gas abundance ratios for CO-to-$H_2O$ and $CO_2$-to-$H_2O$, estimated from relative gas production rates, are now $Q_{CO}/Q_{H2O}$=0.55 and $Q_{CO2}/Q_{H2O}$=0.23, and the ratio of CO-to-$CO_2$ is ~2.5. These are typical numbers found for long period solar system comets (Harrington-Pinto+2022, Cordiner+2025, Lisse+2025b). The total [CO+$CO_2$]-to-$H_2O$ ratio, 0.78, is very high (A'Hearn+2012), arguing that the gas is being produced from CO-rich ice phases as well as $H_2O$-impurity phases, since the carrying capacity for water ice interstitial impurities is <25%. The new organics gas abundance (assuming a g-factor of 3x$10^{-4}$) vs. water is also high at $Q_{Organics}/Q_{H2O}$= 0.14.

This change from pre-perihelion observations makes sense because in August-2025 3I's behavior was dominated by large icy dust grain emission, with the icy grains too cold to sublimate anything more volatile than $CO_2$ fully. By December-2025, though, the ISO had spent ~3.5 months inside the Solar System's ice line, and all of the cometary constituents, not just the highly volatile $CO_2$ and CO ice portions, was active. I.e., bulk matrix comet material was evaporating in December, releasing everything the comet contains.

The change in behavior can be explained by the thermal wave perturbation produced by 3I's relatively close passage by the Sun (perihelion on 30-Oct-2025 at 1.35au) having reached below any alteration depth (~10m, Carpenter1987; Quirico+2023; Maggiolo+2020, 2025) produced by cosmic rays during 3I's Myrs to Gyrs-long passage through the ISM. Some tension exists because the gas production of 3I had been shown to be dominated by icy coma dust grains in SPHEREx August-2025 pre-perihelion imaging (Lisse+2025a,b), and these "grains" would have had to have been >10m thick to preserve deep unprocessed material. This would imply a huge amount of coma dust mass, because these same large boulders have to provide enough surface area to make the coma ~100x brighter than the nucleus in reflected sunlight. On the other hand, we know the coma grains must be large (mm to dm, at least) because there is no evidence for a radiation pressure dominated anti-solar dust tail. We also know that they still contain some ice, because a broad $H_2O$ ice absorption feature is still present in 3I's spectrum (Fig. 1), and the 0.75 – 1.5 m scattered light continuum is significantly bluer than SPHEREx measurements of zodiacal dust scattering, which is dominated by old, ice-removed comet grains (Nesvorny+ 2010).

Another explanation for the observed changes is that the coma ices able to sublimate at low temperature in August were CO-and-organics-poor but $CO_2$ rich. By contrast, the water ice fraction now boiling-off contains the bulk of the nucleus' organics and CO material. We know that the surface temperature of the coma "grains" is now ~270K from modeling the 3.5-5.0um thermal emission continuum. The nature of 3I's scattered continuum emission has also fundamentally changed, from an ice-dominated reflectance spectrum similar to cliff-type KBOs seen in August-2025 (Lisse+2025a) into that of low albedo dust with spectral signature dominated by bluish light scattering. This is consistent with amorphous carbon- and olivine-rich dust which has lost almost all its ice mantling, as expected for cometary material warmed to >200 K for more than a few days in the vacuum of interplanetary space (Lisse+2021, 2022).

**6. Future Work.** The spectrophotometric measurements reported here are from a preliminary analysis. A more full-up treatment will be produced before 3I/ATLAS passes through SPHEREx's planned survey pattern again in April 2026.

# 8. Figure

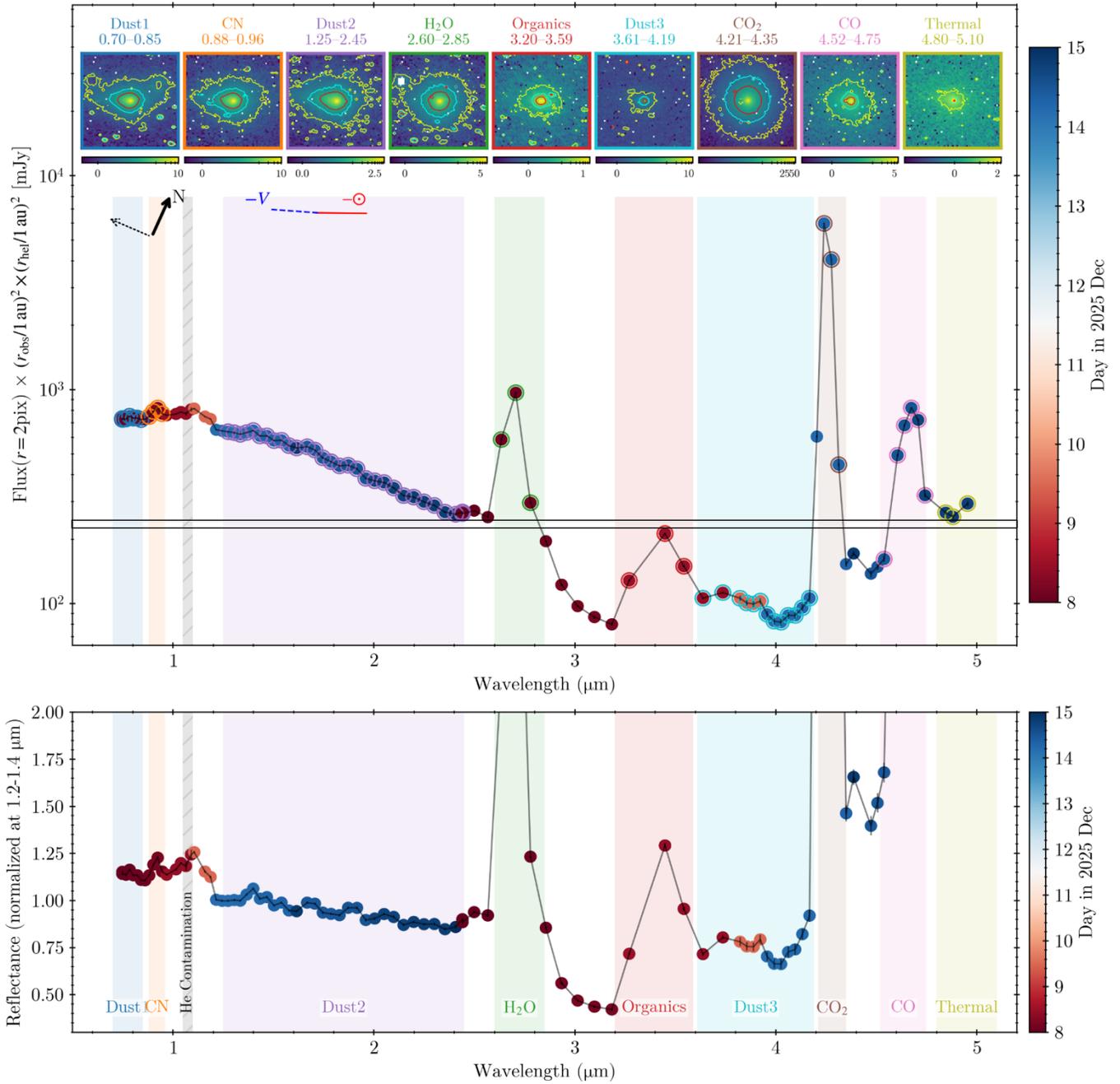

**Figure 1.** (**Top Insets**) SPHEREx 0.75-5.0μm imaging of 3I/ATLAS taken on 08-to-15-Dec-2025. **The 6.3'x6.3' cutout contours are 5,20,&50 times the background noise; color bars are in MJy/sr. The celestial North-and-East directions are shown (black-arrows), along with the anti-velocity (dashed-blue) and anti-sun (solid-red) directions.** Unlike the August-2025

SPHEREx results, all major constituents were resolved vs the SPHEREx 6.2"x6.2" pixel scale. The dust morphology and C-H emission was found to be pear-shaped with a slight anti-tail. **All other** gas comae morphologies were round, arguing for nucleus or near-nucleus gas production. (**Top Plot**) **December-2025 aggregate Flux-Spectrum for 3I/ATLAS,** with **nominal 1-σ uncertainty error bars derived from each pointing's pixel variance.** The spectrophotometry (**symbols**), calculated using a 2-pixel radius aperture, **is colored by observation date (right color bar) and** has been corrected to $r_h$=Delta=1au. Note that the spectrophotometry around 1.083um is highly contaminated by terrestrial exospheric emission. The major gas species spectral emissions [CN(0.93um), $H_2O$(2.7-2.8um), C-H(3.2-3.6um), $CO_2$(4.2-4.3 um), and CO(4.7-4.8 um)] have been color coded for ease of identification. These species are all typical of Solar System comets (Bockelée-Morvan&Biver2017), as is the mixed silicates+amorphous carbon refractory dust fraction dominating the continuum spectrum. The dust spectroscopy can be described by a scattered sunlight + thermal emission model. (**Bottom Plot**) **December-2025 aggregate Reflectance-Spectrum=(Flux-Spectrum/$F_{Sun}$) for 3I/ATLAS. The scattered light continuum, which was approximately neutral in August, is now substantially bluened, suggesting the presence of abundant greenish olivine, some residual ice, and/or a mixture of fine absorbing amorphous carbon+large silicate grains.** Ruled out are particles small enough to bluen via Rayleigh scattering, as these would also be subject to strong solar radiation-pressure effects and would form an anti-solar tail, which is not detected.